\documentclass[%
 reprint,
superscriptaddress,
 amsmath,amssymb,
 aps,
]{revtex4-2}
\usepackage{soul}
\usepackage[english]{babel}
\usepackage{graphicx}
\usepackage{dcolumn}
\usepackage{bm}

\usepackage[version=3]{mhchem} 
\usepackage{siunitx}

\begin{document}

\preprint{APS/123-QED}

\title{Time-resolved Charge Detection in Transition Metal Dichalcogenide Quantum Dots 
}

\author{Markus Niese}
\email{mniese@phys.ethz.ch}
\affiliation{%
 Laboratory for Solid State Physics, ETH Zürich, CH-8093 Zürich, Switzerland
}%
\author{Michele Masseroni}
\affiliation{%
 Laboratory for Solid State Physics, ETH Zürich, CH-8093 Zürich, Switzerland
}%
\author{Clara Scherm}
\affiliation{%
 Laboratory for Solid State Physics, ETH Zürich, CH-8093 Zürich, Switzerland
}%
\author{Christoph Adam}
\affiliation{%
 Laboratory for Solid State Physics, ETH Zürich, CH-8093 Zürich, Switzerland
}%
\author{Max J. Ruckriegel}
\affiliation{%
 Laboratory for Solid State Physics, ETH Zürich, CH-8093 Zürich, Switzerland
}%
\author{Artem O. Denisov}
\affiliation{%
 Laboratory for Solid State Physics, ETH Zürich, CH-8093 Zürich, Switzerland
}%
\author{Jonas D. Gerber}
\affiliation{%
 Laboratory for Solid State Physics, ETH Zürich, CH-8093 Zürich, Switzerland
}%
\author{Lara Ostertag}
\affiliation{%
 Laboratory for Solid State Physics, ETH Zürich, CH-8093 Zürich, Switzerland
}%
\author{Jessica Richter}
\affiliation{%
 Laboratory for Solid State Physics, ETH Zürich, CH-8093 Zürich, Switzerland
}%

\author{Kenji Watanabe}
\affiliation{%
 Research Center for Electronic and Optical Materials, National Institute for Materials Science, 1-1 Namiki, Tsukuba 305-0044, Japan
}%
\author{Takashi Taniguchi}
\affiliation{%
 Research Center for Materials Nanoarchitectonics, National Institute for Materials Science,  1-1 Namiki, Tsukuba 305-0044, Japan
}%
\author{Thomas Ihn}%

\author{Klaus Ensslin}%
\affiliation{%
 Laboratory for Solid State Physics, ETH Zürich, CH-8093 Zürich, Switzerland
}%
\affiliation{%
 Quantum Center, ETH Zürich, CH-8093 Zürich, Switzerland
}%



\date{\today}

\begin{abstract}

We investigate electronic transport through gate-defined quantum dots in molybdenum disulfide (\ce{MoS2}) using an integrated charge detector. We observe a crossover from two weakly coupled single dots to a strongly coupled double quantum dot. In the regime of extremely weak dot-lead coupling, where the direct transport current is below the detection limit, we measure the dot occupation via charge detection and access the few-electron regime. Due to the large band gap of \ce{MoS2}, tunneling rates can be sufficiently suppressed to resolve individual tunneling events. These results establish a platform for single-shot spin- and valley-to-charge conversion and highlight the potential of transition-metal dichalcogenide quantum dots for quantum information applications.
\end{abstract}

\maketitle

Transition metal dichalcogenides (TMDs) offer unique properties for quantum dot physics. The heavy atoms in TMDs give rise to strong spin–orbit coupling~\cite{kormanyosMonolayerMoS22013} (up to $\SI{150}{\milli\electronvolt}$), and their large direct bandgap ($1-2\unit{\electronvolt}$) provides ideal conditions for creating gate-defined quantum dots (QDs).
Combined with their valley degree of freedom~\cite{xiaoCoupledSpinValley2012}, TMD quantum devices are a promising yet unexplored platform for applications in quantum information. Many-electron QDs based on TMDs~\cite{wangElectricalControlCharged2018} and charge detection of incidental QDs in TMDs~\cite{boddison-chouinardChargeDetectionUsing2022} have been reported. However, crucial components for semiconductor qubit operation~\cite{burkardSemiconductorSpinQubits2023}, such as single carrier QDs combined with charge detectors, have not yet been implemented. Time-resolved charge transport, which is necessary for the state readout~\cite{elzermanSingleshotReadoutIndividual2004a}, has not yet been realized with TMD QDs. These steps are necessary to utilize the potential of QDs in TMDs.

There have been several realizations of QDs in both \ce{MoS2} \cite{zhangElectrotunableArtificialMolecules2017, pisoniGatetunableQuantumDot2018, krishnanSpinValleyLockingInGap2023} and \ce{WSe2} \cite{davariGateDefinedAccumulationModeQuantum2020, boddison-chouinardGatecontrolledQuantumDots2021}. Coulomb blockade was demonstrated in all of these devices, but in the many-carrier regime, where it is difficult to understand the spectrum. One of the challenges of operating QDs in TMDs in the few-electron regime is the relatively large effective mass leading to small energy level spacings~\cite{yuPhaseTransitionEffective2015}. Furthermore, they have high impurity densities exceeding $10^{12}-10^{13}\unit{\per\square\centi\meter}$~\cite{edelbergApproachingIntrinsicLimit2019}. Even for advanced growth techniques where impurity densities as low as $10^{10}\unit{\per\square\centi\meter}$ have been reported~\cite{liuTwoStepFluxSynthesis2023}, the mobility is still below \SI{1000}{\square\centi\meter\per\volt\per\second}, which is one to two orders of magnitude lower than in graphene~\cite{deanBoronNitrideSubstrates2010}. Consequently, for bulk samples, a metal-insulator transition is observed at relatively high carrier densities of around $1.7\times 10^{-12} $ \SI{}{\per\square\centi\meter} \cite{masseroniEvidenceCoulombGap2023}, where for lower densities, the charge carriers are localized. This strong localization may completely pinch off tunneling barriers and suppress the direct current through the dot long before the few carrier regime in the dot is reached.

Here, we overcome this limitation by fabricating a second dot (charge detector) capacitively coupled to the investigated dot (signal dot). This allows us to detect the charge occupation in the signal dot even when its source--drain current is below the detection limit. This technique has been pioneered in GaAs \cite{fieldMeasurementsCoulombBlockade1993a} and was used in various other materials before including other systems based on van der Waals heterostructures \cite{kurzmann_charge_2019}.

We detect the addition or removal of charge carriers in the signal dot by observing a shift in the resonance of the detector dot. We show that we can detect charge occupation far beyond the point where current through the signal dot becomes suppressed, and use it to detect both single dots and a double dot. The high tunability of the dots combined with the time-resolved charge detection makes this platform promising for future quantum devices.

\begin{figure*}
    \centering
    \includegraphics[width=\linewidth]{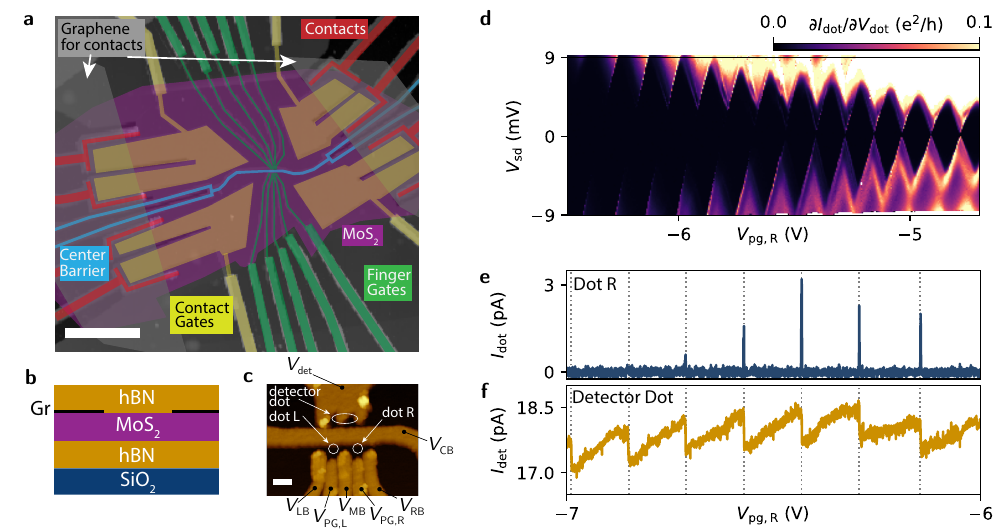}
    \caption{(a) False-color AFM image of the entire device. The \ce{MoS2} flake is purple and stretches horizontally over the whole device. The scale bar is \SI{5}{\micro\meter}. (b) Cross-section of the stack with the four layers of \ce{MoS2} encapsulated by two flakes of hBN and partially in contact with two distinct monolayer graphene sheets forming the electrical contacts to the \ce{MoS2}. (c) Zoom-in image of the gate structure close to the dots. The center barrier connected to $\mathrm{V_{CB}}$ divides the device in two halves. On the top half, the detector dot side, $\mathrm{V_{Det}}$ controls the detector dot. On the bottom half, the signal dot side, the five gates connected to $\mathrm{V_{LB}}$, $\mathrm{V_{PG,L}}$, $\mathrm{V_{MB}}$, $\mathrm{V_{PG,R}}$, $\mathrm{V_{RB}}$ control dot L and dot R. The scale bar is \SI{100}{\nano\meter}. (d) Coulomb diamond measurement of the dot sitting between $\mathrm{V_{MB}}$, $\mathrm{V_{PG,R}}$ and $\mathrm{V_{RB}}$. (e) Current through the signal dot and (f) through the detector dot, measured simultaneously for a source drain bias $V_\mathrm{sd}=\SI{100}{\micro\volt}$. The steps in the detector signal always occur at the same gate voltage as the resonances of the signal dot, but can even be observed when the resonances on the signal side are suppressed.}
    \label{fig1}
\end{figure*}

Figure \ref{fig1}a shows a false-color AFM image of the device. The gates are fabricated on top of the van der Waals heterostructure. Figure \ref{fig1}b shows the composition of the stack, which has been fabricated using the dry-transfer method \cite{deanBoronNitrideSubstrates2010}. It is made, from bottom to top, of a bottom hBN (\SI{40}{\nano\meter} thick), four layers of \ce{MoS2}, a top hBN (\SI{20}{\nano\meter} thick), and metallic top gates made of Ti/Au (\SI{3}{\nano\meter} /\SI{20}{\nano\meter}). The stack is placed on a \ce{Si} / \ce{SiO2} chip with a \SI{280}{\nano\meter} thick oxide. The p-doped Si substrate is used as the back gate.
We use graphene contacts to make an electrical connection to the \ce{MoS2} layer \cite{cuiMultiterminalTransportMeasurements2015}. Therefore, single layer graphene is stacked directly above the \ce{MoS2} flake. In Figure \ref{fig1}a, the two graphene flakes are sketched in gray, spanning vertically on both sides of the device and overlapping with the \ce{MoS2}, shown in purple. The graphene and the \ce{MoS2} flakes overlap only in the contact area. By dry etching with \ce{CHF3} and \ce{O2}, each overlapping area is divided into four individual contacts of around \SI{5}{\micro\square\meter}, thereby enabling four-terminal conductance measurements. The graphene itself is then contacted with edge contacts with Cr/Au (\SI{3}{\nano\meter} /\SI{70}{\nano\meter}) \cite{wangOneDimensionalElectricalContact2013a}, shown in red in Fig. \ref{fig1}a. We place additional metallic gates on top of the contacts (shown in yellow in \ref{fig1}a) to accumulate carriers in the contact area independent of the carrier density in the \ce{MoS2}, which is controlled by the silicon back gate. Using that procedure, we reach contact resistances of around \SI{1}{\kilo\ohm}.

The active area of the \ce{MoS2} can be electronically divided into two parts by applying a negative voltage to the \SI{50}{\nano\meter} wide center barrier ($V_\mathrm{{CB}}$), marked in blue in Fig. \ref{fig1}a. To independently form dots, we tune $V_\mathrm{{CB}}$ such that the region below the central barrier is completely pinched-off. By doing so, we can independently form dots in the two parts. On both sides, we have finger gates marked in green. Figure \ref{fig1}c shows an AFM image of the dot area. On the lower half of the image, the signal dot side, there are five individual gates. Three of the gates act as barriers, labeled left $V_\mathrm{{LB}}$, middle $V_\mathrm{{MB}}$, and right $V_\mathrm{{RB}}$. They are located \SI{100}{\nano\meter} from the center barrier. The other two gates are interleaved between the barrier gates and positioned an additional \SI{50}{\nano\meter} further from the center barrier edge. They act as plunger gates $V_\mathrm{{PG,L}}$ and $V_\mathrm{{PG,R}}$. This gate geometry can be used to form either two individual decoupled dots (labeled dot L and dot R in Fig.~\ref{fig1}c) or a double dot.
The upper half of Fig. \ref{fig1}c shows the detector side. There is one large gate connected to $V_\mathrm{{det}}$ that can form a detector dot as labeled in Fig. \ref{fig1}c.

We first focus on measurements of single dots performed in a dilution refrigerator at a base temperature of \SI{60}{\milli\kelvin}. We apply a positive voltage to the back gate to accumulate electrons in the \ce{MoS2} layer. Closing the barriers by applying negative voltages to $V_\mathrm{{MB}}$ and $V_\mathrm{{RB}}$ leads to a regime in which we observe Coulomb blockade over a wide range of plunger gate voltages $V_\mathrm{{PG,R}}$. Figure \ref{fig1}d shows a measurement of the differential conductance through dot R as a function of the plunger gate voltage $V_\mathrm{{PG,R}}$ and the source-drain voltage bias $V_\textnormal{SD}$. We observe Coulomb diamonds indicating a charging energy of around \SI{3}{\milli\electronvolt} for plunger gate voltages $V_\mathrm{{PG,R}}$ between \SI{-4.5}{\volt} and \SI{-5}{\volt}. Using a capacitor model, this corresponds to a dot radius of around \SI{110}{\nano\meter}. With a typical effective mass $m_e^* \approx 0.4-0.8\quad m_e$ for \ce{MoS2} \cite{pisoniAbsenceInterlayerTunnel2019,yuPhaseTransitionEffective2015, bielTunabilityEffectiveMasses2015} this corresponds to a quantization energy between \SI{15}{\micro\electronvolt} and \SI{30}{\micro\electronvolt}, comparable to the thermal smearing $4k_BT = \SI{20}{\micro\electronvolt}$. For more negative voltages of $V_\mathrm{{PG,R}}$, the current drops below the detection limit, even for biases up to several \unit{\milli\volt}. By depleting the dot more and more, we reduce the dot--lead coupling so much that the current becomes undetectable. This behavior of the dot is consistent with previous results \cite{songGateDefinedQuantum2015, wangElectricalControlCharged2018}, where direct transport through the dot could only be observed at a high number of carriers in the dot. From these measurements, it is apparent that we will be unable to investigate the few-electron regime by measuring the current through the dot.

We can reach the few-electron regime by using the second side of our device for charge detection. By bringing the one big gate defining the detector region (connected to $V_\mathrm{det}$ in Fig. \ref{fig1}c) close to pinch-off, a narrow channel is formed where individual conductance resonances are observed. The localized states responsible for these resonances are separated from the signal dot by the depleted area below the \SI{50}{\nano\meter} wide center barrier. Thus, we can use any of the resonances of such a detector dot for capacitive charge sensing. Figure \ref{fig1}e shows the current $I_\textnormal{dot}$ through the dot R measured at low $V_\textnormal{SD}$ fading out for $V_\textnormal{pg,R}<\SI{-6.8}{\volt}$. Whenever there is a resonance in the current $I_\textnormal{dot}$ through the dot, the current $I_\textnormal{det}$ of the charge detector, shown in Fig.~\ref{fig1}f, jumps. These jumps persist even when the direct current is completely suppressed due to opaque barriers. 

In order to stay in a sensitive regime for large ranges of plunger gate voltages, we correct for the small influence of $V_\mathrm{PG,R}$ on the detector resonance by adjusting $V_\mathrm{dot}$ while sweeping $V_\mathrm{PG,R}$. 
The same procedure is also performed when changing barrier gates, each with individual, slightly different, adjustment factors.

By decreasing the voltages $V_\mathrm{LB}$ and $V_\textnormal{PG,L}$ and simultaneously setting $V_{\mathrm{RB}}$ and $V_{\mathrm{PG,R}}$ to \SI{0}{V} we now populate dot L.
This dot is defined by the barriers $V_\mathrm{{LB}}$ and $V_\mathrm{{MB}}$ and its chemical potential can be tuned by $V_{\mathrm{PG,L}}$. The direct transport signal from dot L is completely suppressed, even for small negative voltages. We assume that this is due to some microscopic differences between dot L and R. However, we can detect the charge state with the detector. To do this, we operate the dot L similarly to dot R.

\begin{figure}
    \centering
    \includegraphics[width=\linewidth]{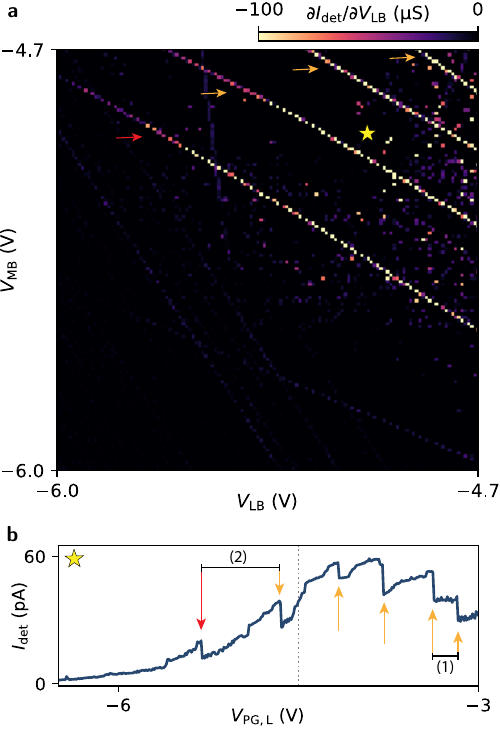}
    \caption{(a) Measurement of the detector current while sweeping the two barriers $\mathrm{V_{MB}}$ and $\mathrm{V_{LB}}$ and keeping $\mathrm{V_{PG,L}}$ constant. Orange arrows point to a set of parallel resonances, with the last one of them marked by the red arrow. The yellow star marks the barrier configuration of the measurement in (b). (b) Detector current when sweeping $\mathrm{V_{PG,L}}$ at fixed barrier voltages. The dotted line indicates the plunger gate voltage at which the measurement in (a) was taken.}
    \label{fig2}
\end{figure}

In order to experimentally establish the few carrier regime, in Figure \ref{fig2}a we analyze a conductivity measurement as a function of the two barrier gate voltages $V_{\mathrm{LB}}$ and $V_{\mathrm{MB}}$ at a constant plunger gate voltage $V_\mathrm{{PG,L}}$. For better visibility of the resonances, we plot the derivative of the detector current with respect to $V_\mathrm{{LB}}$. Steps in the measured $I_\mathrm{{det}}$ now appear as spikes in $\partial I_\mathrm{det}/\partial V_\mathrm{LB} $ in Fig. \ref{fig2}a. We can shift the dot between the barriers. When making $V_\mathrm{LB}$ more negative, the dot moves closer to the middle barrier, thus moving the resonance to less negative voltages of $V_\mathrm{MB}$. This causes the diagonal lines whose slope is given by the ratio of the lever arms between the dot and the two gates.
The spacing between the resonance lines increases with increasingly negative barrier gate voltages. We attribute this to the size of the dot getting smaller which leads to both a larger charging and a larger confinement energy. 
The last detectable resonance is marked by the red arrow. 

We now continue to examine the detector signal for the same sequence of resonances by varying the plunger gate voltage at constant barrier gate voltages. Fig. \ref{fig2}b shows this for the barrier gate voltages marked by the yellow star
in Figure \ref{fig2}a. The dotted line in Fig.~\ref{fig2}b indicates the voltage of $V_\mathrm{{PG,L}}$ at which the map in Fig. \ref{fig2}a was taken. In Fig.~\ref{fig2}b we observe a sequence of steps, each marked with an arrow. The separation between the charging steps increases for more negative $V_\mathrm{{PG,L}}$, ranging from \SI{200}{\milli\volt} (marked by interval (1)) to \SI{660}{\milli\volt} (marked by interval (2)). We note that the linear correction applied to $V_\mathrm{dot}$ no longer holds in this regime. This leads to the detector current level not being constant during the measurement. However, the steps can still be clearly identified.

Assuming the lever arm to be constant over the last few resonances yields an increase in charging energy from \SI{8}{\milli\electronvolt} for interval (1) to \SI{26}{\milli\electronvolt} for interval (2). The significant increase in charging energy, compared to the consistent spacing previously observed, suggests that the confinement of the dot undergoes a drastic change in this regime. This is consistent with the observation of an increasing resonance spacing, followed by the termination of the resonance sequence, as shown in Fig. \ref{fig2}a. Such an increase in the separation between resonances is characteristic of the few-electron regime~\cite{kouwenhovenFewelectronQuantumDots2001},  and suggests that these features correspond to the first charge transitions of the quantum dot.

\begin{figure}
    \centering
    \includegraphics[width=\linewidth]{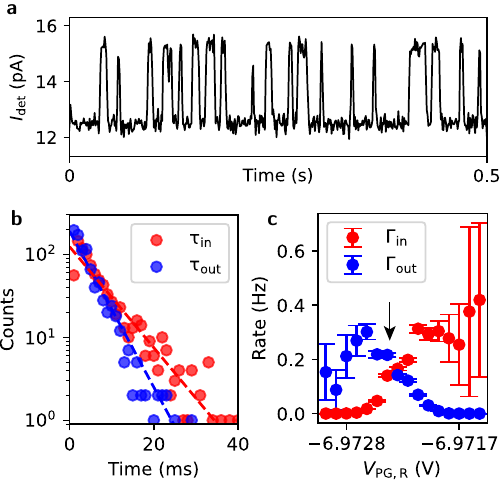}
    \caption{(a) Time trace of $I_\mathrm{det}$ showing different current levels for the electron being in and out of the dot. (b) Histogram of tunneling in and tunneling out events at a certain $V_\mathrm{{PG,R}}$ (marked by the arrow in (c) and fits that show the exponential distribution). (c) Tunneling in ($\mathrm{\Gamma_{in}}$) and tunneling out rates ($\mathrm{\Gamma_{out}}$) as a function of the plunger gate voltage $V_\mathrm{{PG,R}}$. The arrow indicates the rates extracted from (b).}
    \label{fig3}
\end{figure}

We proceed to investigate time-resolved electron-tunneling. To do so, we select a charge transition in the few-carrier regime of dot R where the tunneling rates are between \SI{100}{\hertz} and \SI{1000}{\hertz}. Figure \ref{fig3}a shows the time trace of an electron tunneling in and out of the dot, leading to clearly distinguishable levels and transitions between them. By digitizing the data, we can extract the waiting times $\mathrm{\tau_{in}}$ for tunneling into the dot and $\mathrm{\tau_{out}}$ for tunneling out of the dot. The histogram of those waiting times is shown in Fig. \ref{fig3}b. By fitting an exponential distribution to the histogram, we can extract the tunneling rates $\mathrm{\Gamma_{in}=1/\tau_{in}}$ and $\mathrm{\Gamma_{out}=1/\tau_{out}}$. Figure \ref{fig3}c shows the dependence of these rates on $V_\mathrm{{PG,R}}$ while moving through the transition from $N$ to $N-1$ charge carriers from less to more negative gate voltages. Good control over the tunneling barriers and thus achievable tunneling rates of around \SI{200}{Hz} allow time-resolved measurements in our device even at zero magnetic field.

\begin{figure}
    \centering
    \includegraphics[width=\linewidth]{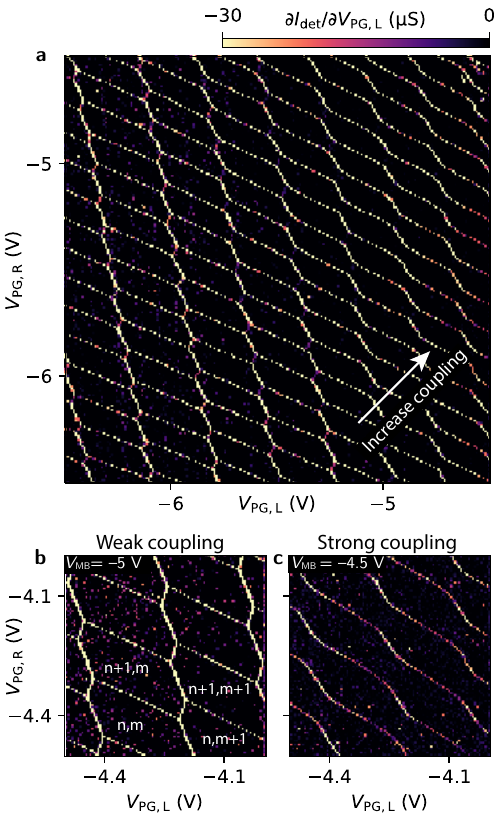}
    \caption{(a) Resonances of a double dot controlled by the plunger gates $V_\mathrm{PG,L}$ and $V_\mathrm{PG,R}$ with constant barrier voltages. (b) Double dot measurement with $V_\mathrm{MB} = $\SI{-5}{\volt}, showing exclusively capacitively coupled dots. We indicate the number of electrons for the dot under $V_\mathrm{PG,R}$ by n and for $V_\mathrm{PG,L}$ by m. (c) The same measurement with $V_\mathrm{MB} = $\SI{-4.5}{\volt}, thus increasing the tunneling coupling, rounding the resonances at their crossings. The color scale is the same for (a), (b) and (c).}
    \label{fig4}
\end{figure}

Double dots can be formed using all the barrier and plunger gates of dot R and dot L as shown in the lower half of Fig. \ref{fig1}c. We control the chemical potentials of the two dots with $V_{PG,L}$ and $V_{PG,R}$, and thus independently of the barriers. With the charge detector being sensitive to both dot L and dot R, we can also use it to measure the charge state of the now formed double dot. Figure \ref{fig4} shows the differential conductance through the detector as a function of both plunger gates. Similar to the single-dot data, we can observe regular dot transitions for a wide range of gate voltages for the double dot. In addition, we can see that by changing the voltage applied to the plunger gate, we can influence the interdot coupling, ranging from purely capacitive coupling in the bottom left corner to an interdot tunneling coupling of approximately $t = \SI{250}{\micro\electronvolt}$ in the top right corner. 
Moreover, we can control the interdot coupling independently of the potentials of the two dots using the middle barrier $V_\mathrm{MB}$. We present the data for the same voltages of $V_\mathrm{PG,L}$ and $V_\mathrm{PG,R}$ with $V_\mathrm{MB} = $\SI{-4.5}{\volt} in Fig. \ref{fig4}b and, $V_\mathrm{MB} = $\SI{-5}{\volt} Fig. \ref{fig4}c, respectively. The capacitive coupling strength between the dots is similar in both measurements with $e^2/C_\mathrm{LR} = \SI{1.8}{\milli\electronvolt}$. The difference in tunneling coupling is clearly visible when comparing the rounded corners of the hexagons. For $V_\mathrm{MB} = $\SI{-4.5}{\volt} (Fig. \ref{fig4}b) we obtain $t \approx \SI{350}{\micro\electronvolt}$ for the tunneling coupling, while for $V_\mathrm{MB} = $\SI{-5}{\volt} (Fig. \ref{fig4}c) $t$ is too small to be extracted from the data.

In conclusion, we have presented single and double dots in \ce{MoS2}. The dots show numerous consecutive Coulomb diamonds, demonstrating the high quality of the dot. The measurements also highlight the limitations of direct transport in this platform due to current suppression. Using a charge detector, we can discover new regimes of quantum dot operation in this material. We showed that steps in the charge detection signal correspond to resonances in direct transport. Importantly, the detection signal can be obtained even when direct transport is suppressed. The good control over the barriers and the potential allow us to do time-resolved measurements and seamlessly transform the system from a single- to a double dot regime, all while using the charge detector to measure the state.
These results demonstrate that charge detection in \ce{MoS2} is possible for single and double dots and can be optimized to perform time-resolved measurements, which is crucial to form spin qubits and investigate their lifetimes in the future.

\begin{acknowledgments}
We are grateful to P. M\"arki, T. B\"ahler, and the FIRST staff for their technical support.
We acknowledge financial support from the European Graphene Flagship, the ERC Synergy Grant Quantropy, and the European Union’s Horizon 2020 research and innovation program under grant agreement number 862660/QUANTUM E LEAPS and NCCR QSIT (Swiss National Science Foundation). K.W. and T.T. acknowledge support from the JSPS KAKENHI (Grant Numbers 21H05233 and 23H02052), the CREST (JPMJCR24A5), JST and World Premier International Research Center Initiative (WPI), MEXT, Japan.
\end{acknowledgments}



\bibliography{CD_QD_TMD_Niese_bib}

\end{document}